# Vavilov-Cherenkov like mechanism of surface wave excitation: details of calculations

V.S.Zuev, A.M.Leontovich, and V.V.Lidsky

The P.N.Lebedev Physical Institute of RAS
119991 Moscow, Russia
vizuev@sci.lebedev.ru
leon@sci.lebedev.ru
vlidsky@mail.ru

This paper contains details of calculation of the surface wave excitation by a fast moving charged particle in a thin uniform metal film surrounded by a dielectic medium. Early on the effect was described in the paper arXiv:0911.0565 03 November 2009. It is shown that the Vavilov-Cherenkov like effect for the surface waves could arise for particles whose velocities are ten- and even hundred-times lower than the corresponding velocities in a uniform medium.

## Механизм Вавилова-Черенкова возбуждения поверхностных волн: детали вычислений

В.С.Зуев, А.М.Леонтович, В.В.Лидский

Физический ин-т им. П.Н.Лебедева РАН
119991 Москва
vizuev@sci.lebedev.ru
leon@sci.lebedev.ru
vlidsky@mail.ru

В данной работе приведены детали расчета эффекта возбуждения поверхностных волн в тонкой однородной пленке металла, окруженной диэлектрической средой, при движении внутри пленки быстрой заряженной частицы. Ранее этот эффект был рассмотрен в работе arXiv:0911.0565 03 November 2009. Показано, что эффект типа Вавилова-Черенкова для поверхностных волн возникает при скоростях частицы в десятки и сотни раз меньших, чем соответствующие скорости в однородной среде.

# Механизм Вавилова-Черенкова возбуждения поверхностных волн: детали вычислений

В.С.Зуев, А.М.Леонтович, В.В.Лидский

Физический ин-т им. П.Н.Лебедева РАН
119991 Москва
vizuev@sci.lebedev.ru
leon@sci.lebedev.ru
vlidsky@mail.ru

В данной работе будут приведены в подробном виде формулы, использованные в статье «Черенковский механизм возбуждения поверхностных волн», arXiv: 0911.0565.

Явление Вавилова-Черенкова – излучение света при движении быстрого электрона в среде наблюдают как в однородных средах /1,2/, так и в неоднородных средах, в таких, как фотонные кристаллы /3/. Условием возникновения этого излучения является наличие в пространстве (однородном или неоднородном) собственных электромагнитных волн с фазовой скоростью меньше, чем скорость пролетающего электрона. Собственные волны с малой фазовой скоростью имеются в нанопленках и в нанонитях из серебра, золота, меди. Это так называемые поверхностные плазмон-поляритоны. Отличие фазовой скорости этих волн от скорости света в вакууме может составлять многие десятки и сотни раз. Это означает, что испускать излучение в виде плазмона может электрон, сравнительно медленный по сравнению с электроном, способным излучать в однородной среде /4/.

Неоднородное пространство с тонкой металлической пленкой имеет в качестве собственных волн симметричный и антисимметричный поверхностные плазмоны без продольной составляющей магнитного поля, так называемые $TM$ - плазмоны, плазмоны поперечно-магнитного типа с магнитным полем в плоскости пленки. Симметрию плазмонов мы определяем по виду магнитного поля. При заданном значении волнового числа частота симметричного плазмона выше частоты антисимметричного плазмона. Поле плазмонов локализовано на пленке, а на удалении от пленки экспоненциально мало. Таких волн в однородном пространстве нет.

Введем декартову систему координат, так чтобы частица двигалась вдоль оси $x$. Оси $x$ и $y$ лежат в плоскости пленки, ось $z$ направим перпендикулярно плоскости пленки. Ток, создаваемый точечной частицей, движущейся со скоростью $v$ вдоль оси $x$, и плотность заряда описываются следующими формулами:

$$j_x = ev\delta(x-vt)\delta(y)\delta(z), \; j_y = j_z = 0, \; \rho = e\delta(x-vt)\delta(y)\delta(z). \tag{1}$$

Вывод начинается с уравнений Максвелла:

$$\nabla \times \vec{H} - \frac{1}{c}\frac{\partial \vec{D}}{\partial t} = \frac{4\pi}{c}\vec{j}, \; \nabla \times \vec{E} + \frac{1}{c}\frac{\partial \vec{B}}{\partial t} = 0,$$
$$\nabla \vec{D} = 4\pi\rho, \; \nabla \vec{B} = 0. \tag{2}$$

Решение уравнений Максвелла следует искать с помощью разложения полей на компоненты Фурье по частоте:

$$\vec{E}(\vec{r},t) = \int_{-\infty}^{\infty} \vec{E}_\omega e^{-i\omega t} d\omega, \; \vec{H}(\vec{r},t) = \int_{-\infty}^{\infty} \vec{H}_\omega e^{-i\omega t} d\omega, \; \delta(x-vt) = \int_{-\infty}^{\infty} \delta_\omega e^{-i\omega t} d\omega. \tag{3}$$

Уравнения Максвелла для компоненты некоторой частоты $\omega$ принимают следующий вид:

$$\nabla \times \vec{H}_\omega + \frac{i\omega}{c}\varepsilon\vec{E}_\omega = -\vec{v}\frac{2e}{cv}e^{i\frac{\omega}{v}x}\delta(y)\delta(z), \tag{4}$$

$$\nabla \times \vec{E}_\omega - \frac{i\omega}{c}\vec{H}_\omega = 0, \tag{5}$$

$$\varepsilon\nabla\vec{E}_\omega = -\frac{2e}{v}e^{i\frac{\omega}{v}x}\delta(y)\delta(z), \tag{6}$$

$$\nabla\vec{H}_\omega = 0. \tag{7}$$

Последнее уравнение следует непосредственно из уравнения $\nabla \times \vec{E}_\omega - i(\omega/c)H_\omega = 0$ в силу того, что для любого $\vec{A}$ имеет место $\nabla \cdot (\nabla \times \vec{A}) = 0$.

Выражение для $\delta_\omega$ получено следующим образом:

$$\delta_\omega = \frac{1}{2\pi}\int\limits_{-\infty}^{\infty}\delta(x-vt)e^{i\omega t}dt = \frac{1}{2\pi v}\int\limits_{-\infty}^{\infty}\delta(x-vt)e^{i\frac{\omega}{v}vt}d(vt) =$$

$$= -\frac{1}{2\pi v}\int\limits_{-\infty}^{\infty}\delta(x-\xi)e^{-i\frac{\omega}{v}(x-\xi)}e^{i\frac{\omega}{v}x}d(x-\xi) = -\frac{1}{2\pi v}e^{i\frac{\omega}{v}x}\int\limits_{-\infty}^{\infty}\delta(\zeta)e^{-i\frac{\omega}{v}\zeta}d\zeta = -\frac{1}{2\pi v}e^{i\frac{\omega}{v}x}$$

Естественно предположить, что у решений системы (4)-(7) зависимость от $x$ имеет вид $e^{i\omega x/v}$. Воспользуемся этим предположением. Итак

$$\vec{E}_\omega(x,y,z) = e^{i\frac{\omega}{v}x}\vec{E}_{\omega k_x}(y,z), \ \vec{H}_\omega(x,y,z) = e^{i\frac{\omega}{v}x}\vec{H}_{\omega k_x}(y,z). \tag{8}$$

Фактически, это есть разложении по $k_x$.

Теперь проделаем разложение по $k_y$.

$$\vec{E}_{\omega k_x}(y,z) = \int\limits_{-\infty}^{\infty}\vec{E}_{\omega k_x k_y}(z)e^{ik_y y}dk_y, \ \vec{H}_{\omega k_x}(y,z) = \int\limits_{-\infty}^{\infty}\vec{H}_{\omega k_x k_y}(z)e^{ik_y y}dk_y.$$

При вычислении производных по $x$ и $y$ от амплитуд $H_{\omega k_x k_y}$ полагаем их равными нулю, так как они от $x$ и $y$ не зависят. В результате получаем

$$\begin{pmatrix} ik_y H_{z\omega k_x k_y} - \partial_z H_{y\omega k_x k_y} \\ \partial_z H_{x\omega k_x k_y} - ik_x H_{z\omega k_x k_y} \\ ik_x H_{y\omega k_x k_y} - ik_y H_{x\omega k_x k_y} \end{pmatrix} + \frac{i\omega}{c}\varepsilon\vec{E}_{\omega k_x k_y} = -\vec{v}\frac{e}{\pi cv}\delta(z). \tag{11}$$

$$\begin{pmatrix} ik_y E_{z\omega k_y} - \partial_z E_{y\omega k_y} \\ \partial_z E_{x\omega k_y} - ik_x E_{z\omega k_y} \\ ik_x E_{y\omega k_y} - ik_y E_{x\omega k_y} \end{pmatrix} - \frac{i\omega}{c}\vec{H}_{\omega k_x k_y} = 0 \tag{12}$$

Уравнение (6) в данном случае не требуется.

Покажем, что эта система (11), (12) распадается на две независимые системы уравнений.

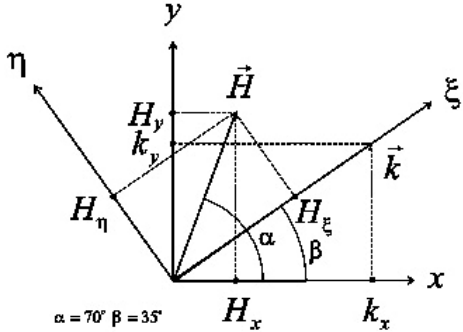

Для этого перейдем в новую систему координат $(\xi,\eta,z)$. Ось $\xi$ совпадает с направлением вектора $\begin{pmatrix} k_x \\ k_y \end{pmatrix}$, ось $\eta$ лежит в плоскости пленки и перпендикулярна к $\xi$, ось $z$ системы $(\xi,\eta,z)$ совпадает с осью $z$ системы $(x,y,z)$.

$$\sin\beta = \frac{k_y}{k}, \ \cos\beta = \frac{k_x}{k}.$$

$$H_\xi = H\cos(\alpha-\beta) \to H_\xi = \frac{1}{k}(k_x H_x + k_y H_y),$$

$$H_\eta = H\sin(\alpha-\beta) \to H_\eta = \frac{1}{k}(k_x H_y - k_y H_x).$$

Для $E_\xi$ и $E_\eta$ берем формулы преобразования, как для $H_i$.

$$H_\xi = \frac{1}{k}(k_x H_x + k_y H_y), \ H_\eta = \frac{1}{k}(k_x H_y - k_y H_x),$$

$$E_\xi = \frac{1}{k}(k_x E_x + k_y E_y), \ \ E_\eta = \frac{1}{k}(k_x E_y - k_y E_x).$$

$$\tag{13}$$

$$H_y = \frac{1}{k}(k_y H_\xi + k_x H_\eta), \ H_x = \frac{1}{k}(k_x H_\xi - k_y H_\eta),$$

$$\tag{14}$$

$$E_y = \frac{1}{k}(k_y E_\xi + k_x E_\eta),\ E_x = \frac{1}{k}(k_x E_\xi - k_y E_\eta).$$

Делаем подстановку в (11).

$$\begin{pmatrix} ikk_y H_z - k_y \partial_z H_\xi - k_x \partial_z H_\eta \\ k_x \partial_z H_\xi - k_y \partial_z H_\eta - ikk_x H_z \\ ik^2 H_\eta \end{pmatrix} = -\frac{i\omega}{c}\varepsilon \begin{pmatrix} (k_x E_\xi - k_y E_\eta) \\ (k_y E_\xi + k_x E_\eta) \\ kE_z \end{pmatrix} - \frac{ek}{\pi c}\begin{pmatrix}1\\0\\0\end{pmatrix}\delta(z).$$

Делаем подстановку в (12).

$$\begin{pmatrix} ikk_y E_z - k_y \partial_z E_\xi - k_x \partial_z E_\eta \\ k_x \partial_z E_\xi - k_y \partial_z E_\eta - ikk_x E_z \\ ik^2 E_\eta \end{pmatrix} = \frac{i\omega}{c}\begin{pmatrix} k_x H_\xi - k_y H_\eta \\ k_y H_\xi + k_x H_\eta \\ kH_z \end{pmatrix}.$$

В результате преобразований получаем тройку уравнений, содержащих только амплитуды $H_\eta, E_\xi, E_z$:

$$-\partial_z H_\eta + \frac{i\omega}{c}\varepsilon E_\xi = -\frac{ek_x}{\pi ck}\delta(z),$$
$$kH_\eta + \frac{\omega}{c}\varepsilon E_z = 0, \qquad (15)$$
$$-\frac{i\omega}{c}H_\eta + \partial_z E_\xi - ikE_z = 0.$$

Эта тройка уравнений описывает поверхностные волны.

Вторая тройка уравнений, уравнений для компонент поля $H_\xi, H_z, E_\eta$:

$$\partial_z H_\xi - ikH_z + \varepsilon i\omega E_\eta = -\frac{e}{\pi}\frac{k_y}{k}\delta(z),$$
$$\partial_z E_\eta + i\omega H_\xi = 0, \qquad (16)$$
$$kE_\eta - \omega H_z = 0.$$

Поле с отличной от нуля $H_z$, следующее из уравнений (16), имеет иную природу, чем поле $TM$-моды (решение 1-й тройки уравнений) с $E_z \neq 0$. Амплитуды поля $H_z$ почти на 3 порядка меньше соответствующих амплитуд поля с $E_z \neq 0$. Практически вся теряемая частицей энергия переходит в излучение $TM$-моды.

Преобразуем систему (15), исключив $E_z$ из 2-го и 3-го уравнений. В результате получим

$$H_\eta = i\frac{c\omega\varepsilon}{c^2 k^2 - \omega^2 \varepsilon}\partial_z E_\xi. \qquad (16)$$

Подставив (16) в 1-е уравнение системы (15), получим

$$\partial_z^2 E_\xi - \left(k^2 - \varepsilon\frac{\omega^2}{c^2}\right)E_\xi = -i\frac{ek_x}{\pi k}\frac{k^2 - \omega^2 \varepsilon/c^2}{\omega\varepsilon}\delta(z). \qquad (17)$$

Это уравнение выполнено как внутри пленки, так и вне ее, причем для поля внутри под $\varepsilon$ следует понимать $\varepsilon_2$ - диэлектрическую проницаемость металла, а для поля выше и ниже пленки — соответственно $\varepsilon_1$. Пространства вне пленки будем отмечать индексами 1 и 3, пространство, занятое пленкой – индексом 2. За пределами пленки среда всюду одинакова.

Введем обозначения:

$$k_{1z} = \sqrt{k^2 - \varepsilon_1 \frac{\omega^2}{c^2}},\ k_{2z} = \sqrt{k^2 - \varepsilon_2 \frac{\omega^2}{c^2}}. \qquad (18)$$

Для полей в пленке и вне пленки получим уравнения:

$$\partial_z^2 E_\xi - k_{2z}^2 E_\xi = -i\frac{ek_x}{\pi k}\frac{k_{2z}^2}{\omega\varepsilon_2}\delta(z), \qquad (19)$$

$$\partial_z^2 E_\xi - k_{1z}^2 E_\xi = 0. \qquad (20)$$

Когда скорость частицы меньше скорости света в среде $v < c/\sqrt{\varepsilon_1}$, то классическое черенковское излучение отсутствует. Однако возможно излучение поверхностных волн. Поверхностная волна распространяется строго по поверхности пленки. Общее решение уравнения (19) является суммой частного решения неоднородного уравнения и общего решения однородного. Частное решение находим методом вариации произвольной постоянной и добавляем общее решение, зависящее от двух постоянных:

$$E_\xi^{(2)} = \frac{e}{2\pi}\frac{k_x}{k}\frac{k_{2z}}{i\omega\varepsilon_2}[\theta(z)e^{-k_{2z}z} + (1-\theta(z))e^{k_{2z}z}] + C_1 e^{k_{2z}z} + C_2 e^{-k_{2z}z}. \tag{21}$$

Здесь $\theta(z) = \int\limits_{-\infty}^{z}\delta(s)ds$ - функция Хевисайда, ступенчатая функция, равная нулю при $z < 0$ и равная единице при $z > 0$.

Решение (21) возникает следующим образом.
Уравнение (19) имеет вид

$$\partial_z^2 E_\xi - p^2 E_\xi = K\delta(z)$$

Решением такого уравнения является сумма частного решения неоднородного уравнения и общего решения однородного. Общее решение однородного уравнения зависит от двух произвольных констант:

$$E_\xi = C_1 e^{pz} + C_2 e^{-pz}.$$

Теперь осталось найти любое решение неоднородного уравнения. Воспользуемся методом вариации произвольной постоянной. Берем любое решение однородного уравнения, произвольную постоянную считаем функцией. Подставляем в уравнение и получаем уравнение меньшего порядка. Возьмем

$$E_\xi = C e^{pz}$$

и подставим в уравнение:

$$C'' + 2pC' = K\delta(z)e^{-pz}.$$

Поскольку $\delta$-функция отличается от нуля только в точке $z = 0$, то правая часть тождественна следующей:

$$C'' + 2pC' = K\delta(z).$$

Интегрируя по $dz$, получим

$$C' + 2pC = K\theta(z) + A.$$

Здесь $A$ - произвольная константа. Действуем тем же способом. Решение однородного уравнения:

$$C = Q e^{-2pz}.$$

Считая $Q$ функцией, находим

$$Q' = [K\theta(z) + A]e^{2pz}.$$

Интегрируем. При $z > 0$:

$$Q = K\frac{1}{2p}(e^{2pz} - 1) + \frac{A}{2p}e^{2pz} + B.$$

При $z < 0$:

$$Q = \frac{A}{2p}e^{2pz} + B.$$

$B$ - некоторая константа. Объединяя, можно записать ответ в виде:

$$Q = K\theta(z)\frac{1}{2p}(e^{2pz} - 1) + \frac{A}{2p}e^{2pz} + B.$$

Откуда для $E_\xi$ находим:

$$E_\xi = K\theta(z)\frac{1}{2p}(e^{pz} - e^{-pz}) + \frac{A}{2p}e^{pz} + Be^{-pz}.$$

Выберем $A$ и $B$ так, чтобы поле убывало при удалении от начала координат как в положительную, так и в отрицательную сторону. Рассматривая область $z < 0$ видим, что $B = 0$. При $z > 0$ приравняем нулю коэффициент при $e^{pz}$:

$$K\frac{1}{2p}+\frac{A}{2p}=0.$$

Откуда $A=-K$. Таким образом:
$$E_\xi = -\frac{K}{2p}[(1-\theta(z))e^{pz}+\theta(z)e^{-pz}].$$

К этому следует прибавить общее решение однородного уравнения. В результате возникает (21).

При подстановке с целью проверки решения (21) в уравнение (19) следует принять во внимание, во-первых, что член $-\delta(z)(e^{k_2 z}-e^{-k_2 z})$ равен нулю при любом $z$, так как при $z=0$ равна нулю скобка $(e^{k_2 z}-e^{-k_2 z})$, а при $z\ne 0$ равна нулю $\delta(z)$. Во-вторых, $\delta(z)(e^{k_2 z}+e^{-k_2 z})=2\delta(z)$. В результате, действительно оказывается, что (21) является решением (19).

Теперь будем рассматривать процедуру вычисления интеграла
$$H(x,y,z,t)=\int_{-\infty}^{\infty}dk_y\int_{-\infty}^{\infty}d\omega H_{\omega k_x k_y}e^{-i\omega(t-x/v)}e^{ik_y y}.$$

Получив зависимость от $\omega$ и $k_y$ амплитуд э.м. поля, можно вычислить сами поля.
$$H_{\omega k_x}(y)=\int_{-\infty}^{\infty}H_{\omega k_x k_y}e^{ik_y y}dk_y$$

Ограничим интервал интегрирования величиной $b$.
$$H_{\omega k_x}(y)=\int_{-b}^{b}H_{\omega k_x k_y}e^{ik_y y}dk_y.$$

Заменим интеграл суммой по формуле трапеции:
$$H_{\omega k_x}(y)=\frac{\Delta}{2}\sum_{n=-N}^{N-1}\{H_{\omega k_x k_y}(n\Delta)e^{iyn\Delta}+H_{\omega k_x k_y}[(n+1)\Delta]e^{iy(n+1)\Delta}\},$$

$$H_{\omega k_x}(y)=\sum_{n=-(N-1)}^{N-1}H_{\omega k_x k_y}(n\Delta)\exp(iyn\Delta)\Delta+\frac{\Delta}{2}\Big(H_{\omega k_x k_y}(-b)\exp(-iyb)+H_{\omega k_x k_y}(b)\exp(iyb)\Big),$$

где $\Delta=\frac{b}{N}$. Здесь $\Delta$ имеет размерность волнового числа, то есть $cm^{-1}$.

Пусть $y_k=y_{\min}+\frac{\pi}{b}k$, где $y_{\min}=-\frac{\pi N}{b}$, $k\in(0,2N)$. Тогда
$$H_{\omega k_x}(y_k)=\sum_{n=-(N-1)}^{N-1}H_{\omega k_x k_y}(n\Delta)e^{i\pi(-N+k)n\frac{\Delta}{b}}\Delta+\frac{\Delta}{2}\Big(H_{\omega k_x k_y}(-b)e^{-i\pi(-N+k)}+H_{\omega k_x k_y}(b)e^{i\pi(-N+k)}\Big),$$
$$n=-N+m.$$

В преобразованном виде
$$H_{\omega k_x}(y_k)=$$
$$=\sum_{m=1}^{2N-1}H_{\omega k_x k_y}(-b+m\Delta)e^{i\pi(-N+k)(-N+m)\frac{1}{N}}\Delta+\frac{\Delta}{2}\Big(H_{\omega k_x k_y}(-b)e^{-i\pi(-N+k)}+H_{\omega k_x k_y}(b)e^{i\pi(-N+k)}\Big)$$
$$H_{\omega k_x}(y_k)=$$
$$=\sum_{m=1}^{2N-1}H_{\omega k_x k_y}(-b+m\Delta)e^{i\pi(N-k-m+km/N)}\Delta+\frac{\Delta}{2}\Big(H_{\omega k_x k_y}(-b)e^{i\pi(N-k)}+H_{\omega k_x k_y}(b)e^{-i\pi(N-k)}\Big)$$
$$H_{\omega k_x}(y_k)=$$
$$=\sum_{m=0}^{2N-1}H_{\omega k_x k_y}(-b+m\Delta)e^{i\pi(N-k-m+km/N)}\Delta-H_{\omega k_x k_y}(-b)e^{-i\pi k}\Delta+$$
$$+\frac{\Delta}{2}\Big(H_{\omega k_x k_y}(-b)e^{-i\pi k}+H_{\omega k_x k_y}(b)e^{i\pi k}\Big)$$

Мы считаем число $N$ четным, при $k$ - целом $e^{-i\pi k} = e^{i\pi k}$.

$$H_{\omega k_x}(y_k) = $$
$$= \sum_{m=0}^{2N-1} H_{\omega k_x k_y}(-b+m\Delta) e^{-i\pi m} e^{i\pi km/N} \Delta e^{-i\pi k} + \frac{\Delta}{2}\left(H_{\omega k_x k_y}(b) - H_{\omega k_x k_y}(-b)\right) e^{-i\pi k}$$

Определим:
$$H_c(-b+m\Delta) = (-1)^m H^*_{\omega k_x k_y}(-b+m\Delta), \; m \in [0, 2N-1].$$

Тогда
$$H_{\omega k_x}(y_k) = \left[\sum_{m=0}^{2N-1} H_c(-b+m\Delta) e^{-i\pi km/N}\right]^* \Delta e^{-i\pi k} + \frac{\Delta}{2}\left(H_{\omega k_x k_y}(b) - H_{\omega k_x k_y}(-b)\right) e^{-i\pi k}.$$

Выражение в квадратных скобках как раз совпадает с формулой функции $fft(H_c)$ как она описана в MATLAB'е.

$$H_{\omega k_x}(y_k) = \{[fft(H_c)]^* + [H_{\omega k_x k_y}(b) - H_{\omega k_x k_y}(-b)]/2\}\Delta e^{-i\pi k}.$$

Теперь надо вычислить распределения полей:
$$H(x,y,t) = \int_{-\infty}^{\infty} H_{\omega k_x}(y) e^{i\omega(t-x/v)} d\omega.$$

Поскольку физические значения $H(x,y,t)$, очевидно, вещественны, то ясно, что $H_{-\omega k_x} = H^*_{\omega k_x}$.

$$H(x,y,t) = \int_0^{\infty} H_{\omega k_x}(y) e^{i(\omega t - \omega x/v)} d\omega + \int_{-\infty}^0 H_{\omega k_x}(y) e^{i(\omega t - \omega x/v)} d\omega =$$
$$= \int_0^{\infty} H_{\omega k_x}(y) e^{i(\omega t - \omega x/v)} d\omega + \int_0^{\infty} H^*_{\omega k_x}(y) e^{-i(\omega t - \omega x/v)} d\omega = 2\operatorname{Re}\int_0^{\infty} H_{\omega k_x}(y) e^{i(\omega t - \omega x/v)} d\omega$$

Пусть существенный вклад вносит область спектра $(\omega_{\min}, \omega_{\max})$. Заменим интеграл суммой:

$$H(x,y,t) = 2\operatorname{Re}\sum_{n=1}^{N}(\omega_{\max} - \omega_{\min})[H_{n-1}(y)e^{i\omega_{n-1}q} + H_n(y)e^{i\omega_n q}]/2N.$$

Здесь $q = t - x/v$, $\omega_n = \omega_{\min} + (\omega_{\max} - \omega_{\min})n/N = \omega_{\min} + \Delta n$.

$$H(x,y,t) = 2\operatorname{Re}\Delta\sum_{n=1}^{N} H_{n-1}(y)e^{i\omega_{n-1}q} + \Delta\operatorname{Re}[H_N(y)e^{i\omega_N q} - H_0(y)e^{i\omega_0 q}] =$$
$$= 2\Delta\operatorname{Re}e^{i\omega_{\min}q}\sum_{n=1}^{N} H_{n-1}(y)e^{i\Delta(n-1)q} + \Delta\operatorname{Re}e^{i\omega_{\min}q}\left(H_N(y)e^{i(\omega_{\max}-\omega_{\min})q} - H_0(y)\right)$$

Пусть $q_k = 2\pi(k-1)/\Delta N$, $k \in [1,N]$. Используем равенство $e^{i2\pi(k-1)} = 1$ при целом $k$.

$$H(x_k, y, t) = 2\Delta\operatorname{Re}e^{i\omega_{\min}q_k}\sum_{n=1}^{N} H_{n-1}(y)e^{i2\pi(k-1)(n-1)/N} + \Delta\operatorname{Re}e^{i\omega_{\min}q_k}[H_N(y) - H_0(y)].$$

Поскольку результат заведомо вещественен, то комплексное сопряжение можно опустить.
На этом исчерпывается рассмотрение деталей вычислений, результаты которых приведены в /4/.